# Azetidinium Lead Iodide for Perovskite Solar Cells

*Samuel R. Pering[a], Wentao Deng[a], Joel R. Troughton[c], Ralf G. Niemann[a], Federico Brivio[a], Peter S. Kubiak[a,b], Florence E. Jeffrey[d], Trystan M. Watson[c], Paul.R. Raithby[a], Andrew L. Johnson[a,d], Simon E. Lewis[a,d], & PJ. Cameron[a,d]*
[a]: *Department of Chemistry, University of Bath, Claverton Down, Bath, BA2 7AY, UK*
[b]: *Centre for Doctoral Training in New and Sustainable Photovoltaics, University of Liverpool*
[c]: *SPECIFIC, Swansea University College of Engineering, Bay Campus, Swansea, SA1 8EN, UK*
[d]: *Centre for Sustainable Chemical Technologies, University of Bath, Claverton Down, Bath, BA2 7AY, UK*

Email: S.R.Pering@bath.ac.uk, P.J.Cameron@bath.ac.uk

Hybrid organic-inorganic perovskites have been established as good candidate materials for emerging photovoltaics, with device efficiencies of over 22 % being reported. There are currently only two organic cations, methylammonium and formamidinium, which produce 3D perovskites with band gaps suitable for photovoltaic devices. Numerous computational studies have identified azetidinium as a potential third cation for synthesizing organic-inorganic perovskites, but to date no experimental reports of azetidinium containing perovskites have been published. Here we prepare azetidinium lead iodide for the first time and show that it is a stable, bright orange material that can be successfully used as the absorber layer in solar cells. We also show that it is possible to make mixed cation devices by adding the azetidinium cation to methylammonium lead iodide. Mixed azetidinium-methylammonium cells show improved performance and reduced hysteresis compared to methylammonium lead iodide cells.

1. Introduction

The amount of research into Organo Lead Halide Perovskites for Perovskite Solar Cells (PSC) has increased rapidly since 2012.[1-3] The benefits of PSC include fabrication using facile solution processing methods,[2,4] and the ability to easily tune properties like band gap and colour.[5,6] Methylammonium lead iodide (MAPI) solar cells have reached efficiencies of over 15 %.[7] MAPI has a bandgap of 1.6 eV,[8] which is higher than the optimum band gap for solar cells of 1.1 – 1.4 eV.[9,10] Band gap engineering is possible by mixing halide ions to form $MAPbI_{3-y}X_y$, where X is either chloride or bromide.[6,8] Alternatively the lead cation can be exchanged for tin. $MASnI_3$ has a lower bandgap of 1.3 V, and $MASnI_3$ PSC can reach efficiencies of 5.7 %.[11] However $MASnI_3$ has been shown to be more unstable and more toxic than $MAPbI_3$.[12,13]

To date, variation of the organic cation has received much less attention compared to variation of the halide and group 14 metal components of the perovskite. This is likely due to the perceived lack of alternatives to methylammonium. Formamidinium (FA) is the only alternative organic cation that has been shown to produce a 3D perovskite. $FAPbI_3$ has a band gap of 1.48 eV, and solar cells have been prepared with efficiencies of up to 16 %.[14,15] An nalternative approach is to replace the organic cation with an inorganic caesium cation. $CsPbI_3$ cells have reached 2.9 % efficiency;[16] $Cs^+$ has also been used as an aditive in MAPI cells, improving both the efficiency and stability.[16,17] Mixed cation perovskites, e.g. containing formamidinium and methylammonium show efficiencies of over 18 %, and a band gap more closely aligned to that of its contact layers.[18] Trication perovskites with the formula $Cs_5(MA_{0.17}FA_{0.83})_{95}Pb(I_{0.83}Br_{0.17})_3$ have shown high efficiencies of 22.1 % and improved stability relative to single cation perovskites.[19,20] Methylammonium and formamidinium are by far the most common cations that are used in high efficiency 3D perovskite solar cells. Organo-lead halide perovskites have been produced using *n*-butylammonium cations, but the larger size of the cation means that a 2D rather than a 3D perovskite is created.[21] Guanidinium lead iodide also forms a 2D perovskite, but the addition of small amounts of guanidinium to MAPI can improve the open-circuit voltage in the resulting devices.[22] 2D perovskites have been investigated as absorber layers in PSC and show enhanced stability compared to devices made with 3D perovskites.[21] 2D materials are very promising, although to date the efficiencies are lower than for devices containing 3D perovskites; 2D perovskites can also require high temperature processing.[23]

Substitution of any of the ions in a 3D perovskite will cause a change in the lattice parameters, and band gap of the material.[24] There are three perovskite phases, the α, β, and γ, as well as a non- perovskite δ phase; which are stable at different temperatures and dictate electronic properties.[25] A tolerance factor, which is calculated based on the size of the ions present, can be used to predict whether or not a 3D perovskite phase will form:

$$t = \frac{(r_A + r_X)}{\sqrt{2}(r_B + r_X)} \tag{1}$$

where $r_A$, $r_B$ and $r_X$ are the ionic radii of the components in the general perovskite formula $ABX_3$. The tolerance factor approach has been used to predict the likely structure of new perovskites, where a value of $t = 0.9 – 1.0$ suggests a cubic perovskite phase will be formed. Compensating for the effect of the halide anions on the radius of the inorganic cation yields a modified factor.[26] Cations with radii that are too large to fit within the cubic perovskite parameters (i.e. $r_A$ gives t > 1) form 2D perovskites.[23] The tolerance factor approach has been used to identify other possible ions that could be used to prepare 3D cubic lead halide perovskites suitable for PSC. Suggested organic cations include azetidinium, $[(CH_2)_3NH_2]^+$, hydrazinium $[H_3N-NH_2]^+$ and guanidinium, $[(NH_2)_3C]^+$.[27,28] Hydrazinium was used recently to improve the efficiency of mixed cation inverted structure perovskite cells.[29] The azetidinium (Az) cation has a computationally derived ionic radius of 250 pm, which lies between the ionic radii of MA (217 pm) and FA (253 pm).[30] A simple tolerance factor calculation yields a *t* value of 0.98, within the region that a perovskite structure could be predicted. Factoring in the effect of the halide ions on the $[PbI_6]^{4-}$ octahedra produces a tolerance factor of 1.03, still within the region where a perovskite could be predicted to form.[31] Several computation papers have now predicted that azetidinium should produce stable lead iodide perovskites. The azetidinium cation has been previously been used in metal-organic perovskites which were not for photovoltaic applications.[32,33] In this paper we demonstrate that azetidinium lead iodide (AzPI) is stable and easy to produce. It is a bright orange solid that can be used to prepare PSC both on its own and when combined in a mixed cation solar cell with MA. We show experimentally for the first time that the Az cation is a good option for the engineering of high-efficiency and stable perovskite solar cells.

## 2. Results & Discussion

In this project we prepared both single crystals and thin films of azetidinium lead iodide (AzPI). The properties of the Az cation are compared to those of the commonly used MA and FA cations in Table 1.

**Table 1.** Comparison of three organic cations for PSC regarding ionic size, Goldschmidt tolerance factor, perovskite structure at RT, dipole moment and chemical structure. Dipole moments were calculated for this study.

| Cation | Methylammonium | Formamidinium | Azetidinium |
|---|---|---|---|
| Effective radius[30] | 217 | 253 | 250 |
| Tolerance Factor[27] | 0.912 | 0.987 | 0.980 |
| RT Structure | tetragonal | hexagonal | unresolved |
| Dipole moment (DFT calc.) | 2.176 D | 0.605 D | 2.519 D |
| Chemical Structure | $H_3C-\overset{+}{N}H_3$ | $H_2N\overset{+}{=}\!\!\diagdown\!\!NH_2$ | azetidinium ring structure |

We found that azetidinium iodide was not soluble in either DMF or DMSO, so a sequential deposition approach was used to produce films of AzPI. A solution of azetidinium iodide in isopropanol was spin-coated on top of a $PbI_2$ film. The $PbI_2$ film rapidly turned a glassy orange colour at room temperature (Figure 1a), the colour did not change with annealing. Powder XRD was performed on both MAPI and AzPI films formed by the two-step deposition route. Major reflections in the AzPI diffractogram were at 11.5 °, 24.9 °, 26.2 ° and 30.1 °, and although there is a feature close to the (002) lead iodide peak at 12.7 °, the diffractogram is clearly different from that of pure PbI$_2$.[2] A comparison of the thin film X-ray diffractograms for MAPI and AzPI shows a large difference in the patterns obtained. The peaks in the MAPI spectrum were more intense and had a narrower peak width, suggesting that the MAPI film is more crystalline than the AzPI film. It is likely that chemical bonding effects play an important role in the interactions between cation and inorganic cage, which results in the formation of a more disordered AzPI phase.

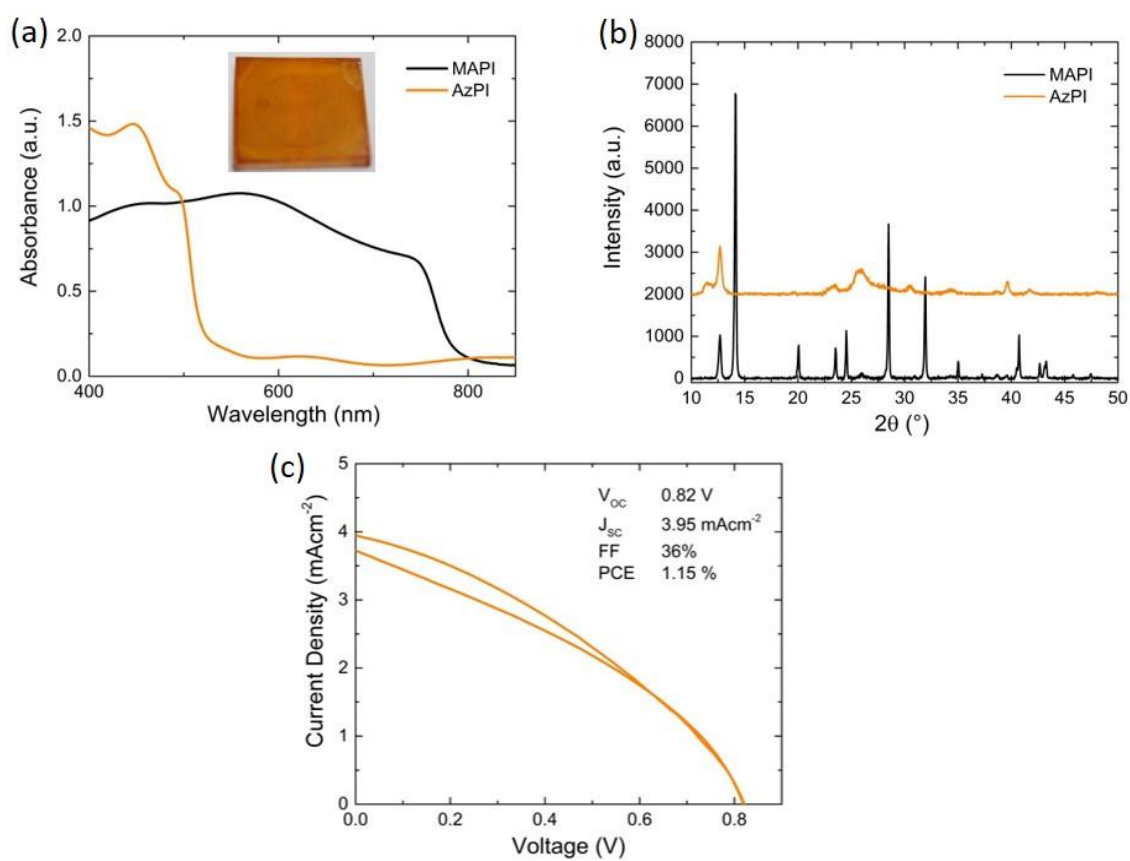

**Figure 1:** Azetidinium Lead Iodide (a) UV/Vis spectroscopy of an AzPI film with MAPI for comparison and (inset) photograph of an AzPI film, (b) XRD patterns of AzPI and MAPI, (c) a J-V curve of an AzPI solar cell with inset cell parameters

Single crystals of AzPI were grown for analysis, but unfortunately a complete description of the atomic positions in AzPI could not be determined by single crystal XRD as all the crystals obtained suffered from severe twinning.

SEM images of the AzPI film (Figure S2) showed that the layer consisted of a large number of small crystals, each around 100 nm in size. There was also a high surface coverage with few noticeable pinholes, which should aid solar cell performance by increasing shunt resistance. A low surface roughness of around 100 nm was observed by AFM (Figure S2).

To estimate the band gap of AzPI, UV/Vis absorbance of the film on glass was measured and compared with a MAPI film (Figure 1a). As expected for an orange film, the absorption onset for AzPI is at shorter wavelengths than for MAPI, with a difference of just over 200 nm. The band gap of AzPI was estimated to be 2.15 eV – significantly higher than the optimum band gap for solar cell materials. Mesoporous PSC were made using the bright orange azetidinium lead iodide layers (Figure 1d) The best cell exhibited an efficiency of just over 1 % (with an average over 8 pixels of 0.96 % and a standard deviation of 0.08), demonstrating that pure AzPI exhibits reasonable efficiencies for a new photovoltaic material in unoptimised solar cells.

Due to the difficulties in resolving the single crystal structure of AzPI it is not possible to say which perovskite phase has been formed. In order to see if AzPI was photoactive, a cyclic voltammogram was measured under chopped illumination. This voltammogram is shown in Figure S3. In order to stabilise the AzPI film, the electrolyte was 0.1 M azetidinium iodide in isopropanol. A positive photocurrent was observed above 0.2 V (versus Ag/AgCl); at 0.2 V the photocurrent switched and became negative. The response suggests that like other organolead halide perovskites, AzPI is ambipolar.[30]

Raman spectroscopy was performed to elicit more information about the possible role of the azetidinium cation in the structure. AzPI is compared to AzI and $Az^+$ in Figure 2 (A full listing of measured modes is in Table S1).[36] The results show a continuous red-shift for most modes in AzPI with respect to AzI. The magnitude of this shift is mostly based upon a chemical Stark shift as well as increased interaction of the $Az^+$ with its direct environment, due to the spatial constraint; thus increasing the interaction between the $Az^+$ cation and the $[PbI_3]_n$ cage. The magnitude of this shift is mostly between 5 to 10 cm$^{-1}$, with few exceptions. Specifically strong shifts can be seen for the ring deformation ($\nu_3$) and the $NH_2$ wagging ($\nu_{13}$). Firstly, the $\nu_3$ mode shows a red-shift of 15 cm$^{-1}$ in the AzPI structure, potentially caused by sterical hindrance of the inorganic cage. Moreover, the $\nu_{13}$ $NH_2$ wagging mode shows a strong red-shift of 61 cm$^{-1}$, which is significantly larger than any other observed shift. The decrease in frequency suggests a weakening of the bond-strength, presumably through a strengthening interaction with the inorganic scaffold. Besides the increased interaction caused by the higher dipole moment (Table S1) and favourable out-of-ring position of hydrogen atoms for bonding with the inorganic scaffold, the $Az^+$ also possesses fewer internal degrees of freedom because of its limited conformational isomerism. One way to compensate for this would be the formation of a lower-dimensional (not 3D) AzPI structure with higher entropy. Strong hydrogen bonding from the organic cation would offer an alternative bonding motif for this structure, as opposed to the three-dimensional I-Pb-I perovskite scaffold. The strong shift of the $NH_2$ wagging mode $\nu_{12}$ indicates strong bonding action from the amine group (e.g. hydrogen bonding) in the orange AzPI phase. The generally stronger shifts in modes associated with the 2C-position and the ambivalent behavior of 1C-positioned modes suggests that the $Az^+$ has a bridging function in the orange phase.

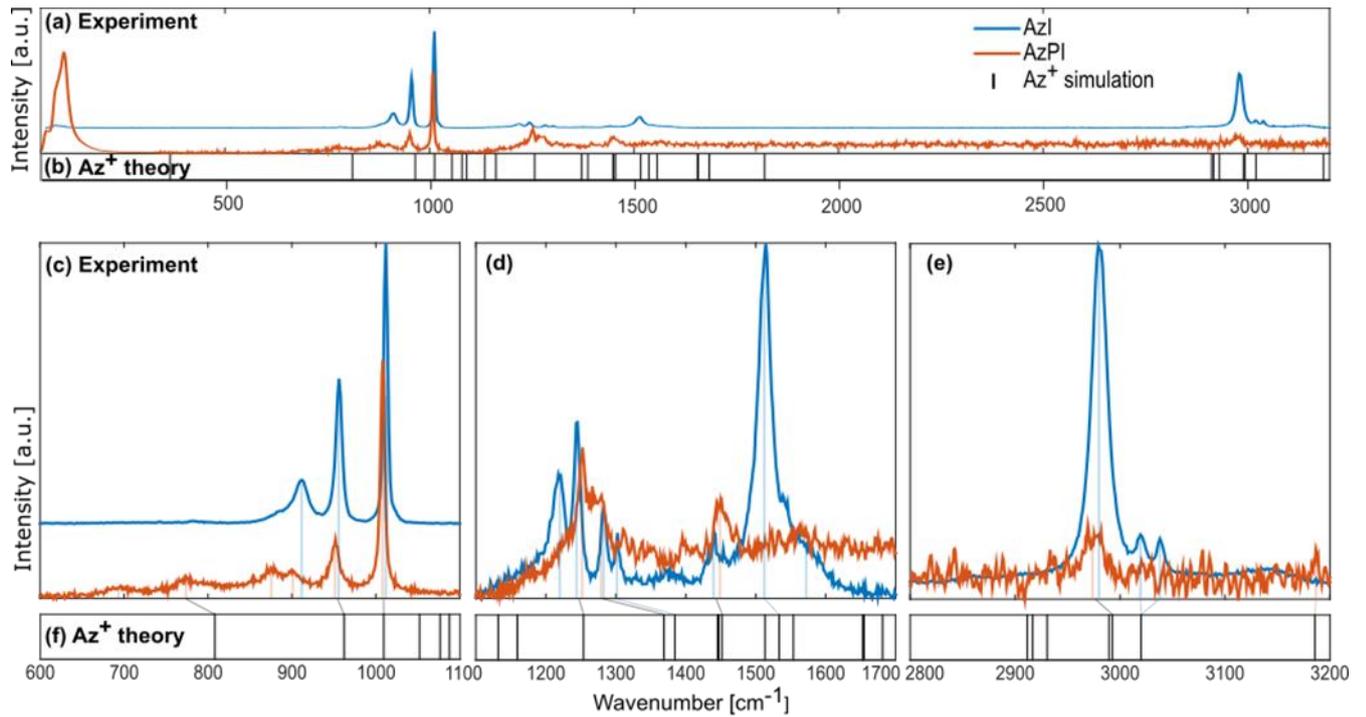

**Figure 2:** (a-b) Full spectra of AzI and AzPI including comparison to $Az^+$ modes (c-e) zoomed region with shifts of assigned peaks against $Az^+$

Both MAPI and FAPI are known to be unstable in even low humidity conditions.[37] To test the stability of AzPI films towards water, both an AzPI film and MAPI film were submerged in water for a few seconds. As might be expected the MAPI film immediately turned yellow on contact with water and part of the film detached into the solution. In contrast the azetidinium film remained fully intact and the bright orange colour was unchanged. To investigate the degree of degradation of each film, thin film XRD was run before and after the dipping experiment. The results are shown in Figure 3a. The AzPI spectrum was largely unaffected by dipping and most importantly there was no increase in the intensity of the $PbI_2$ reflection at 12 °, showing that, unlike MAPI and FAPI, the photoactive phase of AzPI is stable even in the presence of extreme amounts of water. In contrast the MAPI had clearly degraded and the $PbI_2$ peak increased in intensity by more than four times after exposure to water. 2D perovskites have generally been shown to have an improved stability to 3D perovskites with exposure to humidity,[34] but in our study we chose a more extreme test.

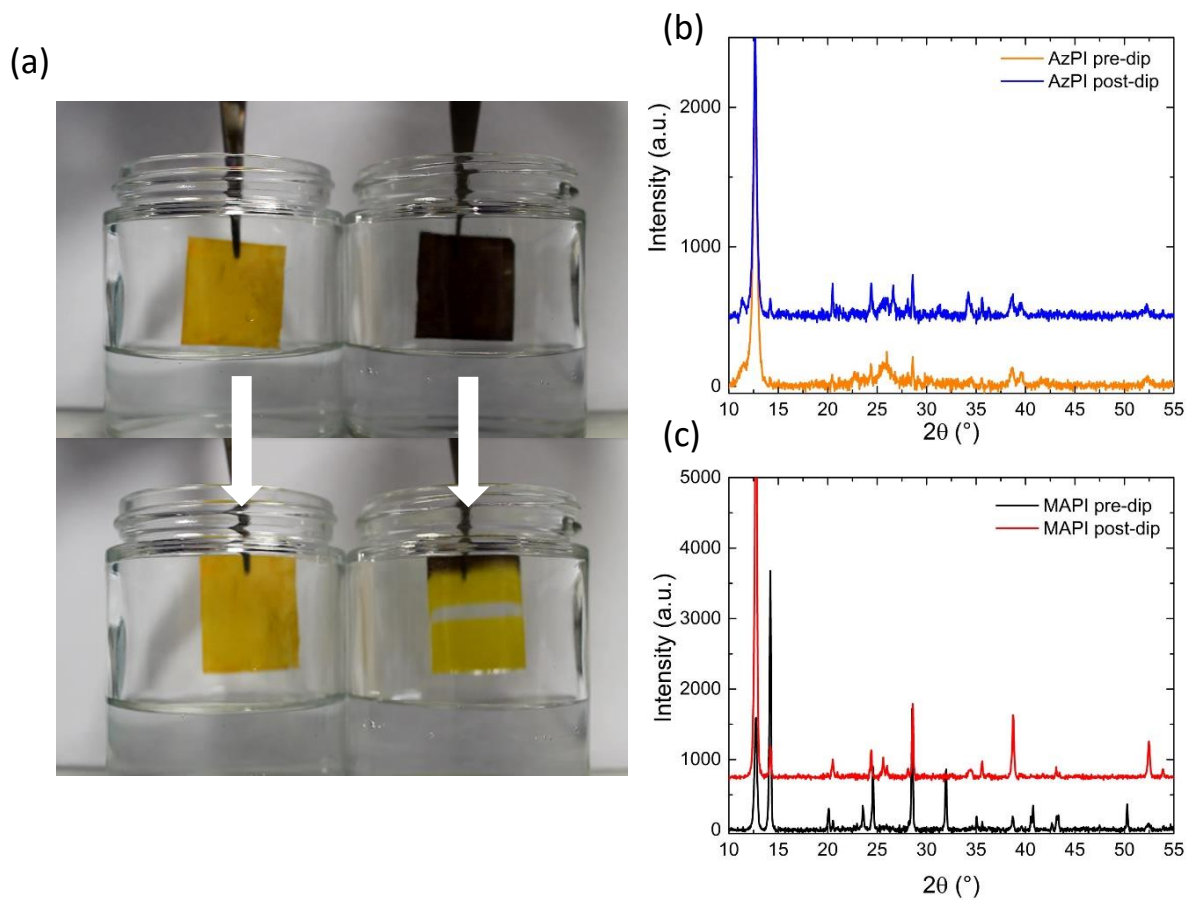

**Figure 3.** The effect of water on AzPI and MAPI: (a) photographs of AzPI (left) and MAPI (right) before (above) and after (below) dipping in water; (b) pre- and post-dip XRD for AzPI, and (c) pre- and post-dip XRD for MAPI

Mixed azetidinium-methylammonium lead iodide PSC were prepared and characterised. Tuning the proportion of different cations in the perovskite is known to alter the band gap and other properties of the material.[20,35] Mixed cation films were prepared by two step deposition; spin-coating methylammonium iodide solutions containing an increasing mole percentage of azetidinium iodide onto a pre-prepared $PbI_2$ film. The resulting films are shown in Figure 4a. The black colour of the MAPI was maintained until 10 mol% AzI was included in the solution. At this concentration the film became visibly lighter in colour. Increasing the mol% of AzI in the solution caused the film to continue lightening towards the orange colour of azetidinium lead iodide. This is evident in the UV/Vis spectra of the films (Figure 4b), in which there is little change in the absorbance until 5 mol% AzI is present in the solution. As the percentage of azetidinium increases further, the absorption onset is slightly blue shifted and there is a reduction in the overall absorbance between 800-600 nm, then an increase in the absorbance at shorter wavelengths.

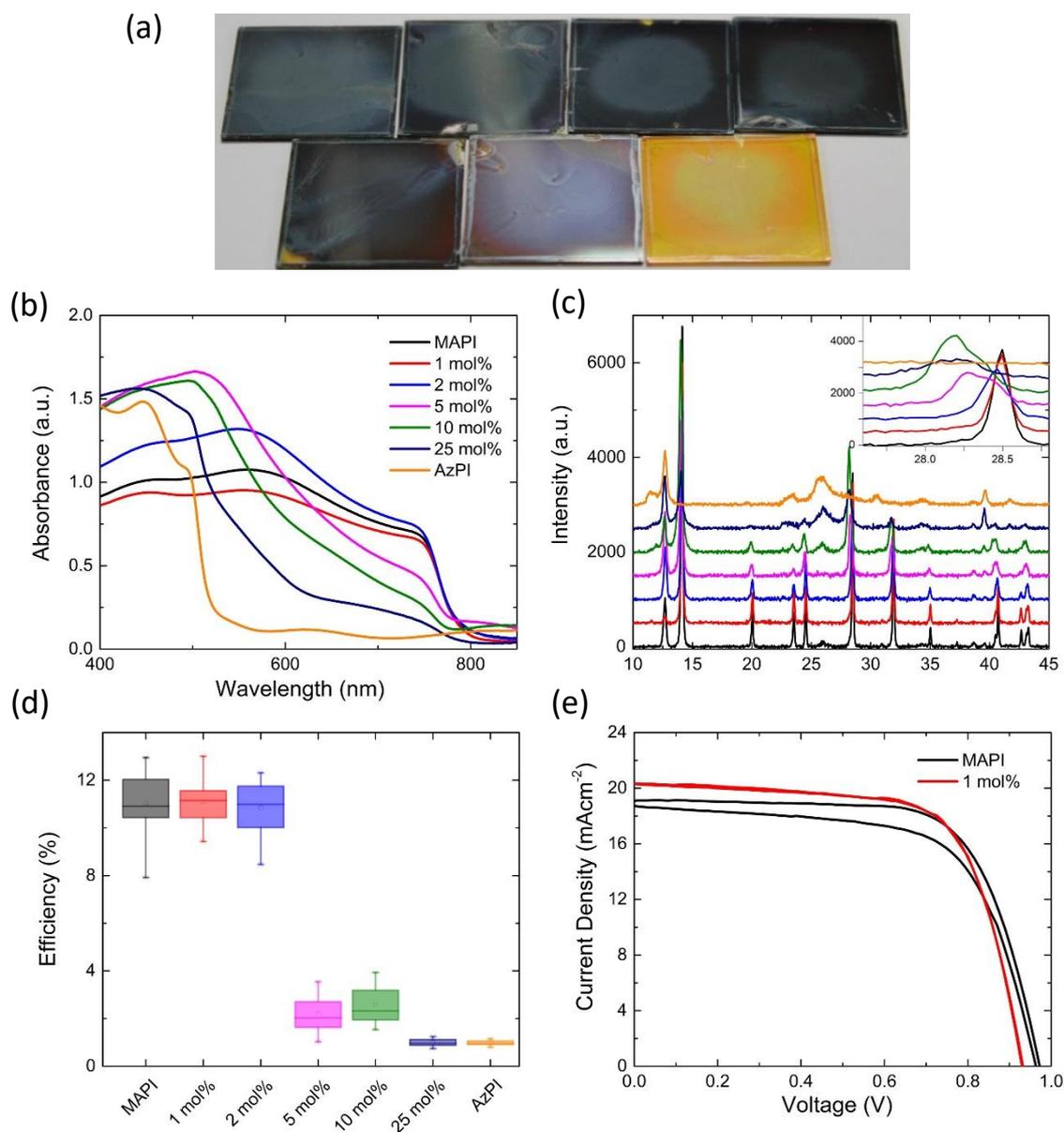

**Figure 4.** A study of the effects of azetidinium on methylammonium lead iodide (a) photographs of the AzMAPI films on glass with different mol% Azetidinium iodide in the spin-coating solution, from left to right: (top row) MAPI, 1 mol% AzI , 2 mol % AzI, 5 mol% AzI; (bottom row) 10 mol% AzI, 25 mol% AzI and AzPI; (b) UV/Vis spectroscopy and (c) X-ray diffraction patterns for the films on glass with inset tracking of the movement of the (2,2,0) peak at 28.5 °; (d) Box-chart for the efficiencies of cells made using AzMAPI and (e) JV curves for the best performing MAPI and AzMAPI cells showing both the forward and reverse scan.

In the thin film X-ray diffractograms there are clear trends that appear with the increasing mole percentage of azetidinium in the spin-coating solution. For compositional ratios of x ≤ 0.02, the diffraction pattern shifts towards smaller angles, caused by an increase in unit cell size. There is a single peak at 28.5 ° without shoulder, suggesting that there is a continuous phase of $Az_xMA_{1-x}PbI_3$. For larger Az ratios 0.05 ≤ x ≤ 0.1, the perovskite reflections indicate splitting into two domains. The primary AzPI peak at 26.2 ° appears at x = 0.1, and upon further increasing the Az ratio to x ≥ 0.25, the intensities of the MAPI phase reflections decrease, while the AzPI reflections become more intense. Taking the MAPI (2,2,0) reflection as an example (inset in Figure 4c), there is a broadening and leftward shift – with the appearance of shoulders in the peak for the 5 mol% AzI sample. The apparent phase separations in the films made with a larger proportion of AzI explains the shape of the UV/Vis spectrum, with separate AzPI phases within the MAPI film causing shoulders in the absorbance, rather than completely shifting the band gap.

For solar cell fabrication $PbI_2$ films were dipped into solutions of mixed AzI and MAI, to allow the organic cation solution to fully penetrate the mesoporous layer (Figure S4). MAPI cells were used as the control. Until recently the highest reported efficiency for a pure $MAPbI_3$ cell was 15%, [2] however using a multi-step Lewis base adduct method, the efficiency can be increased to 19.7 %. [38] The highest efficiency perovskite devices are mixed cation and contain MA, FA and Cs. We chose to work with lower efficiency $MAPbI_3$ cells to allow us to fully characterise any changes introduced by the Az cations. The cells were made by a standard two-step deposition method and were not optimised for efficiency.

**Table 2.** Photovoltaic parameters for MAPI and 1 mol% Azetidinium solar cells (averaged over **15 pixels**)

|  | $V_{OC}$ (mV) | $J_{SC}$ (mAcm$^{-2}$) | FF (%) | Champion Cell (%) | Efficiency (%) |
| --- | --- | --- | --- | --- | --- |
| MAPI | 948 ± 13 | 18.6 ± 0.70 | 62 ± 5.1 | 12.9 | 11.0 ± 1.35 |
| Az$_{0.01}$MA$_{0.99}$PI | 925 ± 16 | 18.1 ± 0.95 | 65 ± 4.0 | 13.0 | 11.1 ± 0.95 |

AzMAPI mixed cation cells deposited from a solution containing more than 5 mol% AzI show a lower efficiency than pure MAPI cells (Figure 4d). This agrees with the UV/Vis spectra, XRD (Figure 3b & 3c) and EQE (Figure S5) measurements which all show phase separation into MAPI and AzPI regions when 5 mol% AzI is present in the precursor solution. A discontinuous phase may hinder charge transfer through the lattice. With lower percentages of azetidinium in solution there is an improvement in the average efficiency of the cells, which is largely due to an improved fill factor (Figure S6). This, as well as a significant reduction in the hysteresis, is evident in the JV curves displayed in Figures 4(e) and S7, where, as in Table 2, the best performing MAPI and AzMAPI pixels are compared. There is a reduction in the standard deviation of the cell efficiency, from 1.35 in the MAPI control to 0.95 in the Az$_{0.01}$MA$_{0.99}$PbI$_3$ set. The stabilised power output of the devices also increased, shown in Figure S8. The best performing cell, from the Az$_{0.01}$MA$_{0.99}$PbI$_3$ set had an efficiency of 13.00 % in the reverse sweep, and 12.98 % in the forward sweep.

3. **Conclusion**

In conclusion, we have synthesised and characterised the new compound, azetidinium lead iodide (AzPI). We have determined that this compound exists as a stable, bright orange film and exhibits some photovoltaic capability, with an optical band gap of 2.15 eV. This tests the application of tolerance factor calculations to these materials, as azetidinium lead iodide is predicted to be within the range where a 3D perovskite should form. Chemical differences in the cation itself (dipole, acidity of the N-H group) are likely to be a key factor that needs to be brought into consideration when making predictions. Azetidinium can be co-deposited with methylammonium to further improve the properties of the MAPI perovskite. These AzMAPI perovskites show an enhanced efficiency and stability compared to pure MAPI, with a reduced hysteresis at low percentages of azetidinium.

## 4. Experimental

*4.1. Dipole calculations*

The calculated dipoles have been obtained using the NWChem code. [39] The initial input obtained by geometrical intuition has been optimized. To express the wavefunction we used the cc-pVTZ basis set provided within the package and the as exchange correlation functional we used the B3lyp. The obtained values are in good agreement with other reports in literature. [40,41]

*4.2. Azetidinium iodide preparation*

5 ml of Azetidine (Alfa Aesar) at 0 °C had 55 mL hydroiodic acid (Sigma) added to it under argon atmosphere **(Caution! Exotherm)**. The ice bath was subsequently removed, and the solution was stirred for one hour. The solution was then left on a rotary evaporator until dry, leaving a bright orange solid. This was washed with diethyl ether to remove the iodine, and recrystallized in isopropanol, leaving white needle-like crystals. The identity of azetidinium iodide (AzI) was confirmed by $^1$H NMR (Figure S1): (300 MHz, D$_2$O, δ): 2.46 (quin, $J$ = 8.29 Hz, 2 H) δ 4.04 (t, $J$ = 8.5 Hz, 4 H)

*4.3. Crystal formation*

0.1 mmol PbI$_2$ and AzI were dissolved in 1 mL *N,N*-dimethylformamide, and single crystals were grown by the solvent evaporation method.

*4.4. Film deposition*

For optical and structural measurements the perovskite films were deposited on to microscope glass following a method by Zheng *et al.*[35] Before film deposition, the substrates were cleaned by sonication in 2% Hellmanex solution in water, followed by deionised water, acetone and isopropanol at 90 °C. Lastly they were treated with UV/Ozone for 20 minutes. 100 µL of a 1 M solution of PbI$_2$ (Sigma-Aldrich) in *N,N*-dimethylformamide was spin-coated at 2000 rpm for 60 seconds, followed immediately by 100 µL of isopropanol spun at the same rate. The resulting PbI$_2$ film was dried at 60 °C for 30 minutes. Solutions containing varying mole percentages of azetidinium iodide compared to methylammonium iodide were prepared in isopropanol, with a concentration of 20 mgml$^{-1}$. 100 µL of these solutions were pipetted onto the PbI$_2$ films, and spun for 60 seconds at 2000 rpm. The perovskite films were annealed at 100 °C for 20 minutes.

*4.5. Solar cell fabrication*

Pre-etched FTO glass (Kintek) was cleaned in 2% Hellmanex solution in water, followed by deionised water, acetone and isopropanol. A compact TiO$_2$ layer was deposited by spray pyrolysis. A hand held atomiser was used to spray a solution of 10 vol% solution of titanium isopropoxide (bisacetylacetonate) (Sigma-Aldrich) in isopropanol onto the substrates, which were kept at 550 °C for the procedure, and sintered for 30 minutes at the same temperature. A mesoporous layer consisting of a 2:7 mixture of 30 NR-D TiO$_2$ paste (Dyesol) in ethanol was spun onto the compact layer with a further 30 minute sintering step at 550 °C. After cooling, to improve conductivity a 0.1 M solution of Li-TFSI (Sigma) solution was spin-coated at 3000 rpm for 10 seconds and the substrates were then re-sintered at 550 °C for 30 minutes. Perovskite deposition was performed in a nitrogen filled glove box. A two-step dip-coating method was used to fabricate the solar cells. 1M PbI$_2$ in DMF was kept at 70 °C for spin-coating. 100 µL of PbI$_2$ solution was spin-coated at 6500 rpm for 30s, then dried at 100 °C for 30 minutes. A 5 minute dipping step in the MAI or mixed MAI/AzI in IPA solution (10 mgml$^{-1}$, AzI fractions in mol% with respect to MAI). The films were annealed at 100 °C for 1 hour.

The hole transport layer solution consisted of 85 mgml$^{-1}$ Spiro-OMeTAD (Ossila) in chlorobenzene, with additives of: 30 µLml$^{-1}$ *t*-butyl pyridine (Sigma), 20 µLml$^{-1}$ of 520 mgmL$^{-1}$ Li-TFSI in acetonitrile and 30 µgmL$^{-1}$ FK209-TFSI solution. This was spin-coated onto the perovskite at 4000 rpm for 30 seconds.

To establish the contacts, 2 mm of perovskite was removed from the centre of the substrate. 100 nm of gold (Kurt J Lesker) was deposited by thermal evaporation.

*4.6. SCXRD, PXRD*

Crystal X-ray diffraction was performed on an Agilent Technologies EOS S2 Supernova, using a Cu X-ray source.

A Bruker axs D8 advance powder x-ray diffractometer with a Cu Kα source and Ge monochromator was used for Powder X-ray diffraction. Measurements were taken from 2θ values of 10 ° to 80 °.

*4.7. UV/vis spectroscopy*

Thin film optical Transmission and Reflectance measurements were performed on a Perkin-Elmer Lambda 750S UV/Vis spectrometer, from 1000 nm to 250 nm. Absorption was calculated as incident light– (transmission + reflectance).

*4.8. Raman spectroscopy*

Raman measurements were performed with a Renishaw in via Reflex microRaman spectrometer equipped with solid state lasers emitting at 514 and 785 nm with a resolution of < 2 cm$^{-1}$. The laser beam was focused with a x50 magnification lens, giving a laser spot size of about 1 µm in diameter. Rayleigh scattering was rejected with a 110 cm$^{-1}$ cutoff dielectric edge filter. The AzI sample was measured with a 514 nm laser and the orange AzPI with the 785 nm laser in order to avoid resonant effects in the sample. All measurements were performed in air and with different laser powers to ensure that the laser probe did not induce damage or changes in the sample

*4.9. Electrochemical measurements*

An Autolab potentiostat/galvanostat was used for solution based electrochemistry, using an Ag/AgCl reference electrode and a platinum counter electrode. 0.1 mol azetidinium iodide in isopropanol was used as the electrolyte. For Mott-Schottky measurements the frequency was 27 Hz and the voltage was swept stepwise from -0.1 to 0.75 V vs Ag/AgCl.

*4.10. J-V curves*

*J-V* curves were measured using a Keithley 2601A potentiostat, under 1 Sun intensity and at AM 1.5. The cell was sweeped at 100 mVs$^{-1}$ from 1.1 V to -0.1 V and back to 1.1 V. The 8 pixels with a 0.1 cm$^2$ active area (obtained using a mask) were measured independently. A Newport Oriel 91150-KG5 reference cell with a KG5 filter was used for instrument calibration.

EQE measurements were taken in 10 nm steps from 380-850 nm.

*4.11. Scanning Electron Microscopy*

SEM images were taken on a JEOL SEM 6480LV, at an acceleration voltage of 10kV.

*4.12 Atomic Force Microscopy*

AFM images were taken on a Nanosurf easyscan 2 FlexAFM system in dynamic mode using a force of 20 nN. A ContAl-G Tip was used for measurements.

**Acknowledgements**
We thank the University of Bath (studentship for SRP), EPSRC Centre for Doctoral Training in New and Sustainable Photovoltaics (studentship for PSK under Grant EP/LO1551X/1) EPSRC Centre for Doctoral Training in Sustainable Chemical Technologies (studentship for FEJ under Grant EP/L016354/1), the EPSRC (PRR under grant EP/K004956 and PJC under grant EP/H026304/1) and the European Union Seventh Framework Programme (RGN [FP7/2007–2013] (DESTINY project)) under grant agreement 316494.

**Supporting Information**

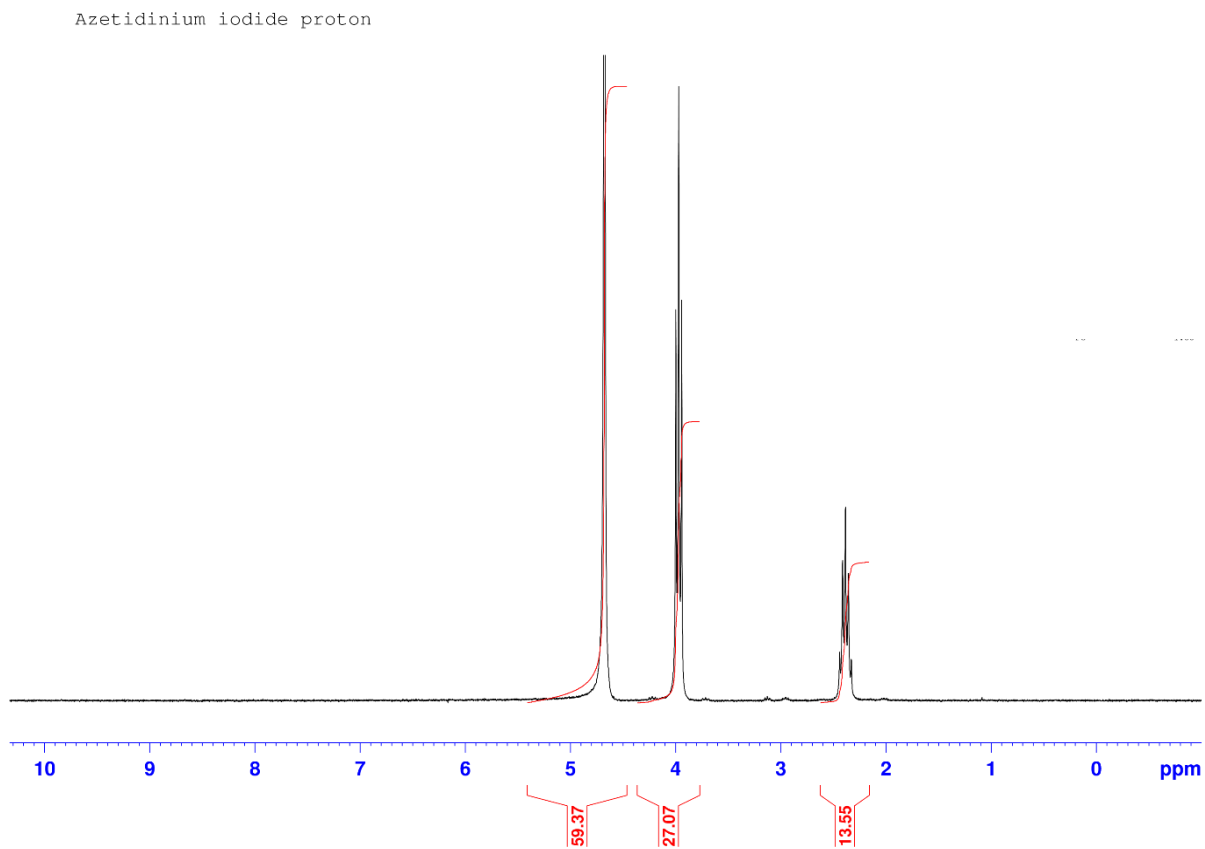

**Figure S1:** Azetidinium Iodide $^1$H NMR Spectrum (Taken in $D_2O$ on a 300 MHz Spectrometer)

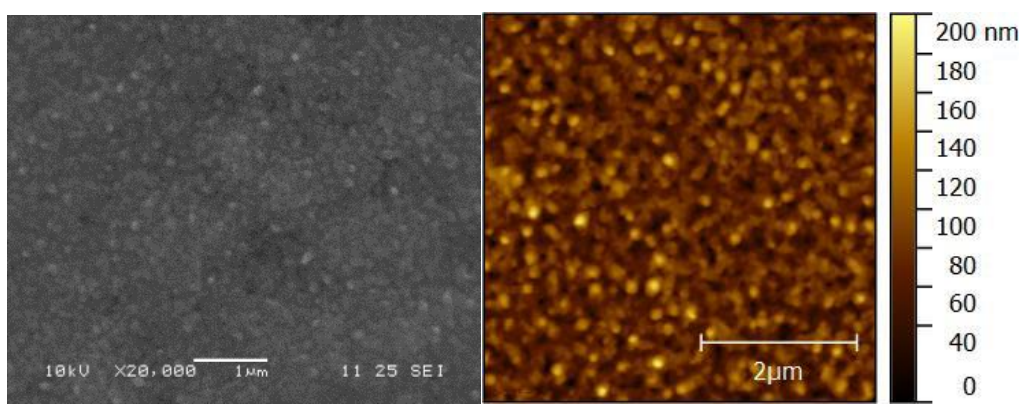

**Figure S2:** SEM image (left) and AFM image (right) of an AzPI film

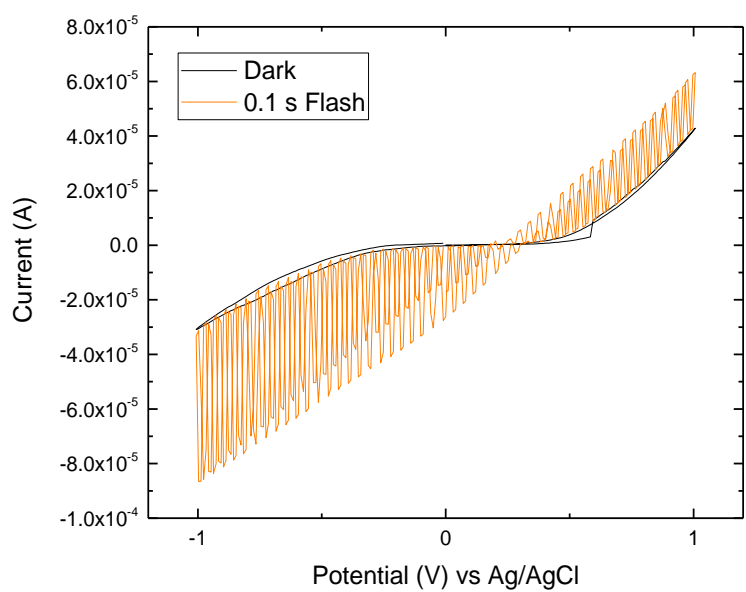

**Figure S3:** Chopped photocurrent measurement of AzPI on FTO, in 0.1 M AzI in IPA electrolyte, with Pt counter electrode and Ag/AgCl reference.

**Table S1.** Comparison of all measured Raman modes of AzPbI$_3$, AzI and Az with provisional peak assignment.

| Az (DFT) | AzI | rel. shift | AzPI | | Assignment |
|---|---|---|---|---|---|
| 361 | | | | $v_1$ | Ring pucker |
| | | | 693 | $v_2$ | N-H bend (in plane) |
| 808 | 785 | ↓ 15 | 770 | $v_3$ | Ring deform |
| | 883 | ↓ 8 | 875 | $v_4$ | 2-CH$_2$ twist |
| | 911 | ↓ 12 | 899 | $v_5$ | 2-CH$_2$ rock |
| 961 | 956 | ↓ 5 | 951 | $v_6$ | 1-CH$_2$ rock |
| 1009 | 1012 | ↓ 4 | 1008 | $v_7$ | 2-CH$_2$ rock |
| | 1217 | | | $v_8$ | 1-CH$_2$ wag |
| 1254 | 1245 | ↑ 7 | 1252 | $v_9$ | 1-CH$_2$ twist |
| 1313 | 1283 | ↓ 6 | 1277 | $v_{10}$ | 2-CH$_2$ twist |
| 1385 | 1302 | | | $v_{11}$ | 1-CH$_2$ twist |
| 1452 | 1455 | | | $v_{12}$ | 1-CH$_2$ wag |
| 1534 | 1511 | ↓ 61 | 1450 | $v_{13}$ | NH$_2$ wag |
| 1655 | 1582 | | | $v_{14}$ | 2-CH$_2$ scissor |
| 2982 | 2980 | ↓ 8 | 2972 | $v_{15}$ | 1-C-H stretch |
| 3020 | 3019 | | | $v_{16}$ | 2-C-H-stretch |

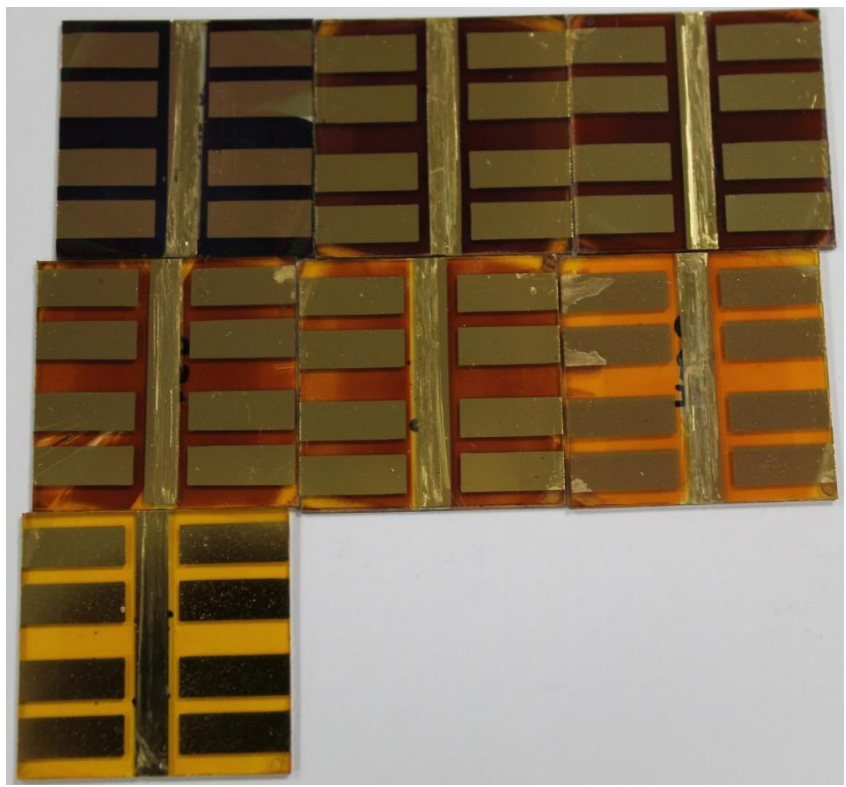

**Fig. S4:** A photograph of the MAPI/AzMAPI solar cell, from left to right: (top) MAPI, A1, A2 (middle) A5, A10, A25 (bottom), AzPI

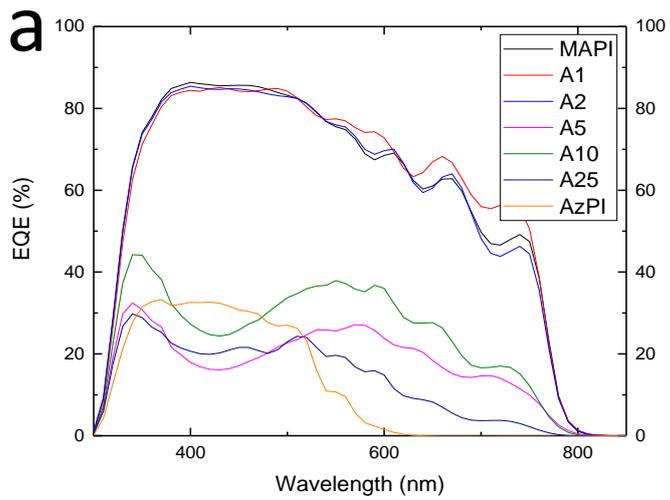

**Fig. S5:** (a) EQE measurements for the best performing pixels for each mixture of MAPI, AzMAPI or AzPI and

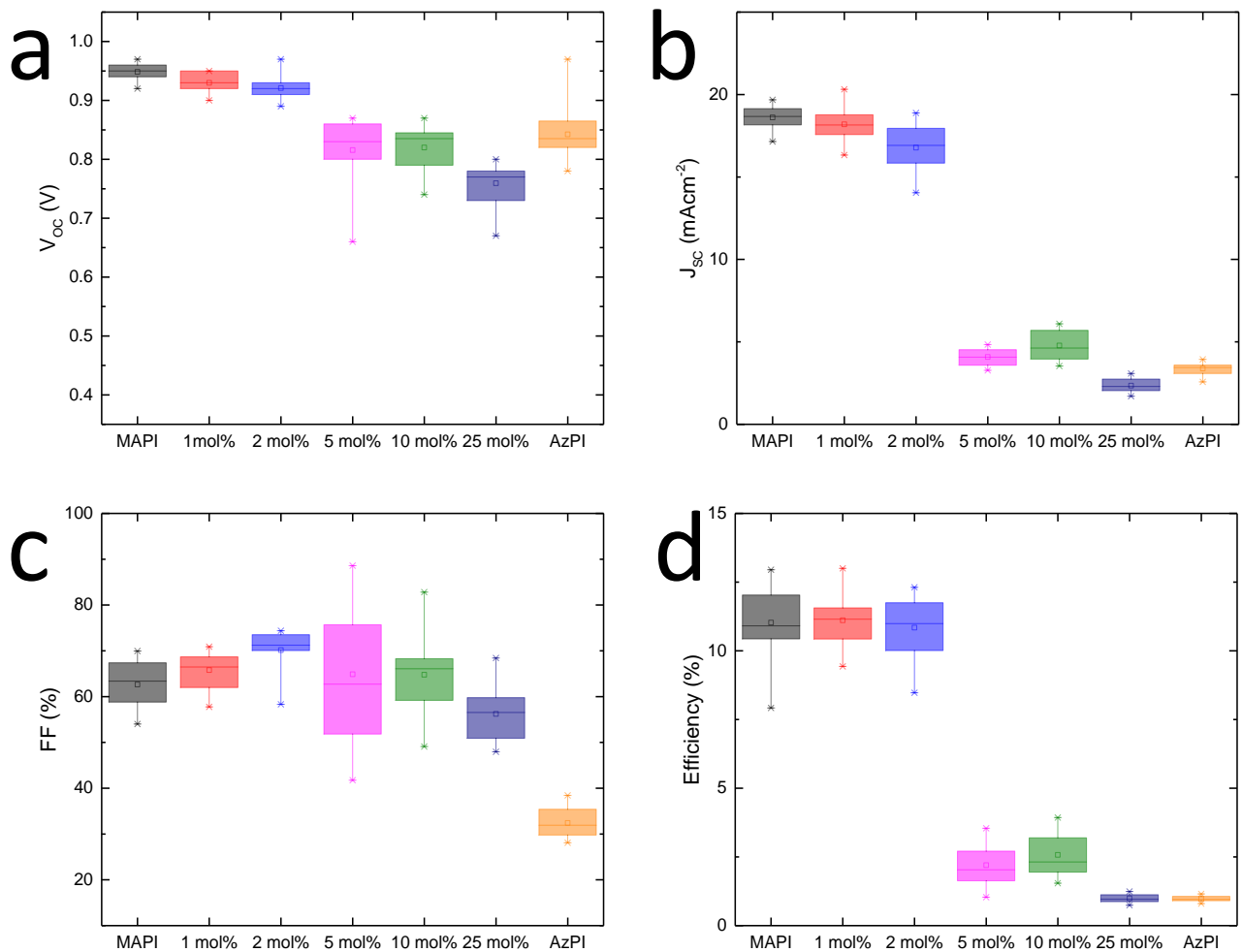

**Fig. S6** Box plots for the cell parameters of AzMAPI cells: (a) $V_{OC}$ (b) $J_{SC}$ (c) Fill Factor and (d) Efficiency

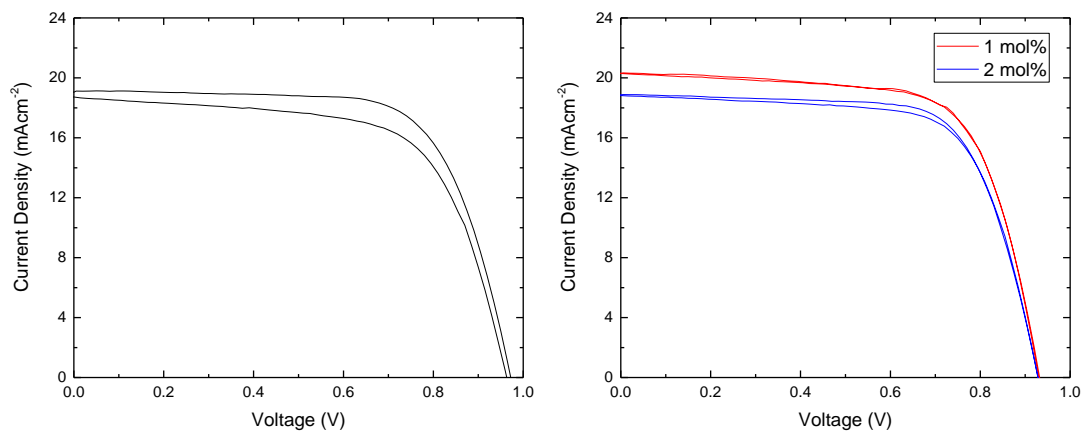

**Figure S7.** A comparison of the hysteresis in the JV curves for MAPI cells (left) and those with azetidinium additives

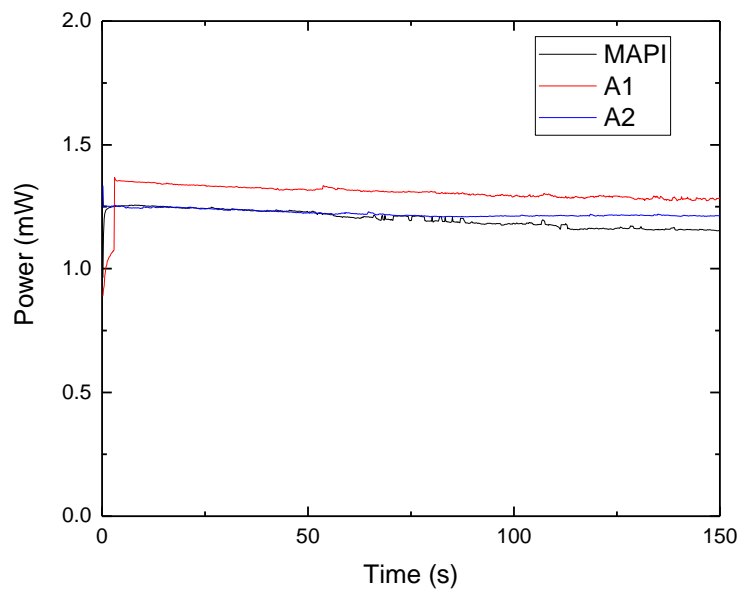

**Figure S8.** Stabilised power output measurements for the best performing AzMAPI and MAPI pixels